\documentclass[dvips]{article}

\def\Journal#1#2#3#4{{#1} {\bf #2}, #3 (#4)}

\def\PRL{\em Phys. Rev. Lett.}

\def\APL{\em Appl. Phys. Lett.}

\def\PRB{{\em Phys. Rev.} B}
\def\SST{\em Semicond. Sci. Technol.}
\def\NAT{\em Nature}
\def\SCI{\em Science}

\def\be{\begin{equation}}
\def\ee{\end{equation}}
\def\bea{\begin{eqnarray}}
\def\eea{\end{eqnarray}}

\usepackage{graphics}
\usepackage{epsfig}

\begin{document}

\title{MODIFICATION OF 1D BALLISTIC TRANSPORT USING AN ATOMIC FORCE MICROSCOPE}

\author{R. CROOK, C.G. SMITH, M.Y. SIMMONS, D.A. RITCHIE}


\maketitle
\begin{center}
Department of Physics, Cavendish Laboratory, Madingley Road, 
\\Cambridge, CB3 OHE, United Kingdom
\end{center}
\begin{abstract}
We have used the scanning charged tip of an Atomic Force Microscope (AFM) to 
produce images of the conductance variation of a quantised 1D ballistic 
channel.               
The channel was formed using electron beam defined 700~nm wide split gate 
surface electrodes over a high mobility GaAs/AlGaAs heterostructure with a two 
dimensional electron gas (2DEG) 98~nm beneath the surface.       
We operate the AFM at 1.5~K and 4.2~K in magnetic fields up to 2~T to observe 
several phenomena.
With a dc voltage on the AFM tip we have produced conduction images of the tip 
potential perturbation, 
as the channel is a sensitive probe of the electrostatic potential.
We have also performed gate sweeps with the tip at a series of points across 
the width of the channel.
The observed structure in transconductance corresponds to the theoretical 
electron density for the first three sub-bands.
When certain gates were biased near pinch off, stable two level switching was 
observed in the images.
We were able to control the state of the switch with the gate bias and tip 
position, and so roughly locate the position of the switching source.
One of the responsible defect systems was located beneath the 2DEG, and due to 
screening of the tip potential, the image reveals the 1D channel.
By asymmetrically biasing the two gate electrodes we have produced images 
showing a total channel movement of 102~nm across the width of the channel, 
but accompanied by 201~nm of movement along the length of the channel which 
is due to imperfections in the surface electrodes and disorder in the doping 
layer.
\end{abstract}

\section{Introduction}
Recent advances in low temperature Scanning Probe Microscopy (SPM) technology
 and the availability of high mobility 2DEGs has made possible the imaging
 of several quantum phenomena.
Images of 2DEG electron compressability in the quantum
 Hall state have been produced~\cite{Tessmer98}, and a single electron transistor (SET)
 has been manufactured on a glass tip to detect static charge~\cite{Yoo97}.
The charged tip AFM has been used to locally perturb the 2DEG electrostatic potential,
 to cause electron backscattering through a 1D channel,
 and so image the ballistic electron flux~\cite{Eriksson96a}.
Numerical calculations of the electric field from a conical tip have shown that the
 half maximum perturbation occurs at a radial distance approximately equal
 to the 2DEG depth~\cite{Eriksson96a}, 
 which limits the spacial resolution of this technique.

\section{Experiment}
In this paper we present images produced by recording the conductance or
 transconductance through a two terminal 1D ballistic channel,
 as a charged AFM tip is scanned over the channel region.
The 1D channel was created in a 2DEG at a GaAs/AlGaAs heterojunction 98~nm
 beneath the surface,
with a $12\times10^{18}$~cm$^{-3}$ Si doped layer from 40~nm to 80~nm
 above the 2DEG.
From Shubnikov-deHaas measurements the carrier concentration was calculated
 as $2.4\times 10^{11}$~cm$^{-2}$ with an electron mean free path of 
 $25~\mu$m.
The 1D channel was defined by locally depleting electrons from the 2DEG
 beneath negatively biased 700$nm wide split gate surface electrodes.
The surface electrodes extend 30$nm above the GaAs surface, 
 and were manufactured using e-beam technology.
Our AFM operates from room temperature to 1.5~K, and uses a piezoresistive
 tip~\cite{PSI} because of the complications of making optical measurements of deflection
 at low temperatures with light sensitive devices.
The AFM operates in the conventional topographic mode to locate the split gate
 region.
During the measurements the AFM force feedback is disconnected and the tip scans
 a constant height above the surface.
\par
We use two experimental configurations to produce images which are referred to as
 conductance images and transconductance images.
In both configurations a lock-in amplifier measures the ac zero phase component
 of the channel drain current.
For conductance images the ac voltage signal was connected to the channel source
 with a dc bias applied to the conductive AFM tip.
For transconductance images the ac signal was connected to the tip with a dc
 bias applied to the channel source.
The channel conductance is a function of $V_{gate}$, 
 so for useful transconductance images we operate within a range of G where 
 $dG/dV_{gate}$ remains almost constant.

\section{Results and discussion}

\subsection{Images of the tip perturbing potential}
\begin{figure}
\begin{center}
\includegraphics[width=11.9cm]{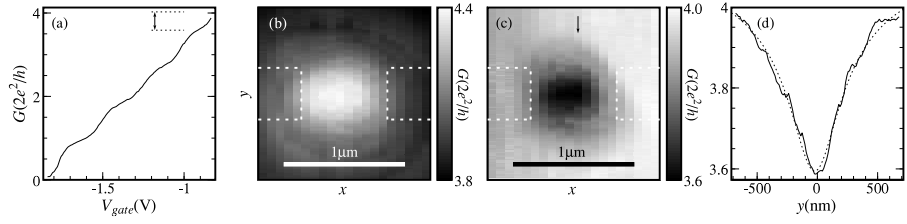}
\end{center}
\caption{Conductance images of the perturbing potential with a tip bias
 of $+2.5$~V in (b) and $-2.5$~V in (c).
A gate sweep is shown in (a) where the tip was off the surface.
A y-direction single sweep indicated by the arrow in (c) is reproduced in (d), 
 where the dotted line shows the result of fitting $V_{tip}/\sqrt{\rho^2+(d+r_{tip})^2}$ to the curve.
\label{figone}}
\end{figure}

In fig.~\ref{figone} we present conductance images where the tip scanned a constant
 60nm above the surface. 
In (b) a $+2.5$~V bias was applied to the tip while in (c) $-2.5$~V was applied,
 with a 0.1~mV ac signal applied to the channel source.
The channel conductance is a sensitive probe of the electrostatic potential~\cite{Field93}, 
 so these images reveal the tip perturbing potential at the 2DEG layer.
Fig.~\ref{figone} (a) shows a gate sweep made with the tip away from the surface, 
 where the dotted lines correspond to the conductance range seen in (c).
If we model the tip as a spherical charge of radius $r_{tip}$ then at a distance $d$ below the 
 tip we have $U\propto V_{tip}/\sqrt{\rho^2+(d+r_{tip})^2}$ where $\rho$ is the radial distance 
 in the 2DEG plane.
Fitting this equation to the y-direction single sweep shown in (d),
 which is indicated by the arrow in conductance image (c), 
 provides a good fit with $d+r_{tip}=230$~nm.
Including a correction for the change in relative permittivity,
 we obtain $r_{tip}=127$~nm which is in agreement with observations from topological images.

\subsection{Measurement of charge density across the channel width}
\begin{figure}
\begin{center}
\includegraphics[width=5.9cm]{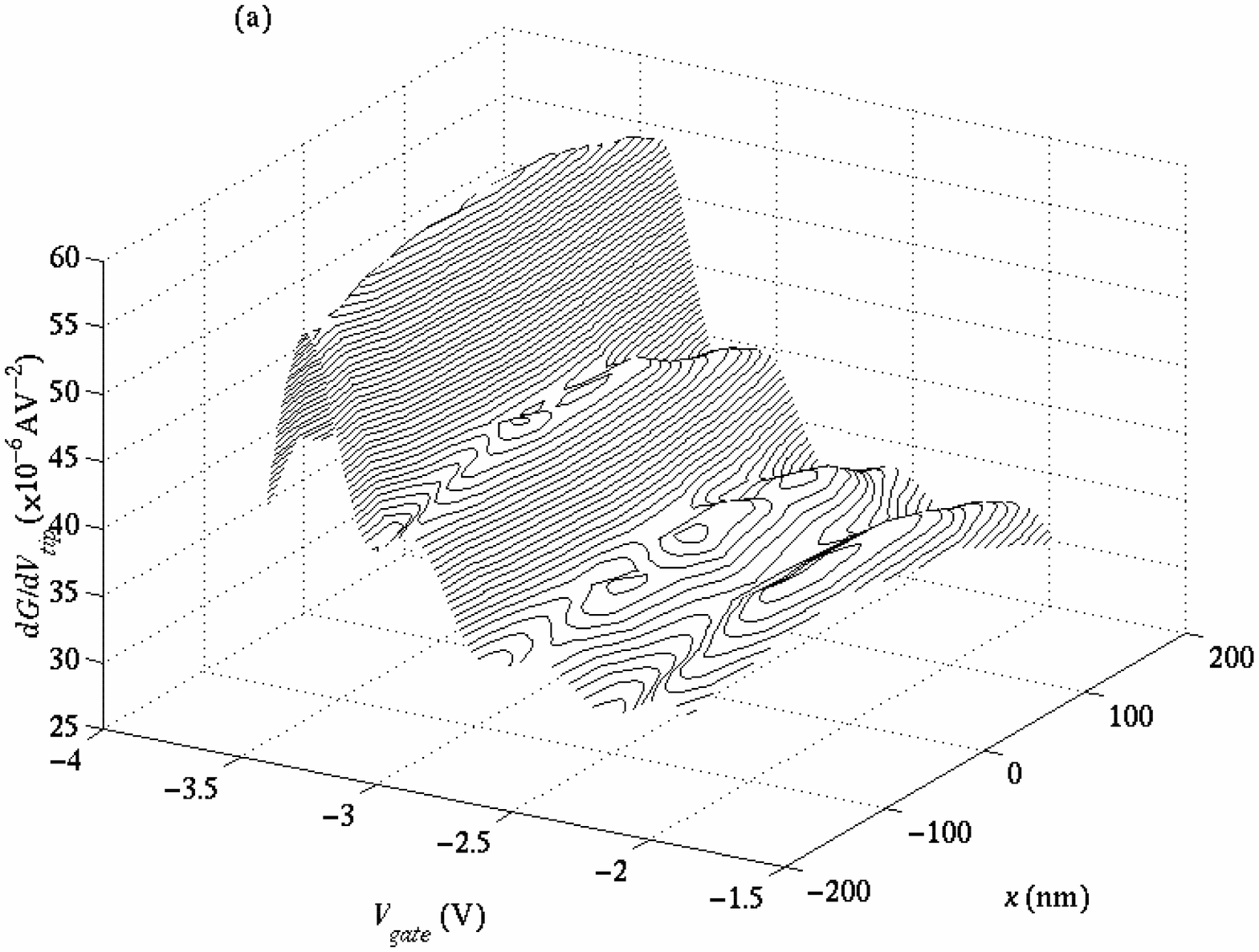}
\includegraphics[width=5.9cm]{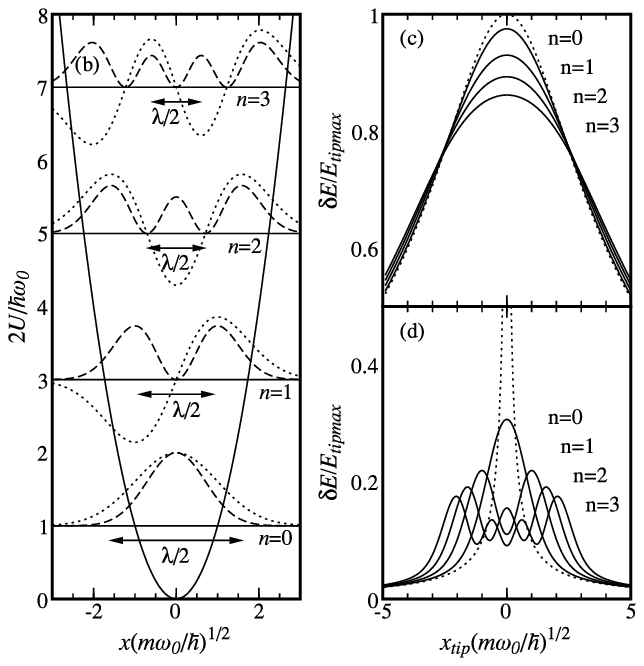}
\end{center}
\caption{A contour plot of transconductance is given in (a) where the gate bias was swept
 at a series of points across the width of a ballistic channel.
The analytic wavefunctions, electron density, energy levels,
 and parabolic confining potential are shown in (b),
 for an infinitely long 1D channel.
The model tip perturbation is shown as a dotted line in (c),
 for a deep 2DEG with $d = 3\left( m\omega_0 / \hbar \right) ^{0.5}$, and in (d),
 for a shallow 2DEG with $d = 0.1\left( m\omega_0 / \hbar \right) ^{0.5} $, 
 where the solid lines are the resulting perturbed energy levels for $n$ = 0, 1, 2, and 3.
\label{figtwo}}
\end{figure}

In fig.~\ref{figtwo} (a) we present a contour plot which relates to the charge density 
 across the width of a quantised 1D channel.
The AFM was positioned at a series of points across the width of the channel (x axis),
 and at each point the transconductance was recorded (z axis) while the
 gate bias was swept from $-2$~V to channel pinch off (y axis).
The series of points passed through the channel centre which was located
 using images like those of fig.~\ref{figone}.
The data of fig.~\ref{figtwo} (a) has been smoothed in the y direction to remove
 small steps caused by switching.
The wavefunctions, and charge density, for the infinite 1D channel with 
 parabolic confinement are well known and shown in fig.~\ref{figtwo} (b),
 where the spacing between the charge density peaks is roughly equal to 
 half the electron wavelength $\lambda$. 
When we introduce the tip perturbing potential to modify the confining potential, 
 the energy levels change by $\delta E$ as shown in (c) for a deep 2DEG,
 and in (d) for a shallow 2DEG, as a 
 function of tip position $x_{tip}$ across the width of the channel.
Note that the peaks in the charge density plots also appear in the same
 positions on the corresponding $\delta E$ plots,
 but convolved with the curve of the perturbing potential.
The significance of the charge density peaks depends on the width of the
 perturbing potential,
 and the model predicts that the $n=1$ peaks will no longer be observed
 when $d > 1.2\left( m\omega_0 / \hbar \right) ^{0.5}$ or $d > 0.3\lambda$,
 although increased broadening for higher sub-bands is still predicted.
\par
An ac signal is applied to the tip which oscillates the energy levels.  
Each energy level determines the gate voltage of the corresponding
 transition between 1D plateaus.
With $V_{gate}$ kept constant,
 the measured transconductance is proportional to a product of the conductance against $V_{gate}$ gradient
 and $\delta E$.
Alternatively, the charge density may modulate the capacitive coupling between the tip and
 the 1DEG to produce a similar response in transconductance.
The distinct peaks in the y-direction of (a) are due the transconductance of 1D quantisation 
 where a peak corresponds to a transition between 1D sub-bands and a trough to a 1D plateau.
On top of the distinct peaks we observe weak oscillations in the x-direction
 where one, two, and three peaks are seen for corresponding transitions
 of up to one, two, and three sub-bands,
 which we interpret as a measure of the charge density across the width
 of a quantised 1D channel.
The $n=1$ peaks are separated by approximately 100~nm,
 giving an electron wavelength larger than the 2DEG Fermi wavelength due to the
 reduced electron effective energy and a finite longitudinal momentum in the channel.

\subsection{Images of switching states}
\begin{figure}
\begin{center}
\includegraphics[width=9cm]{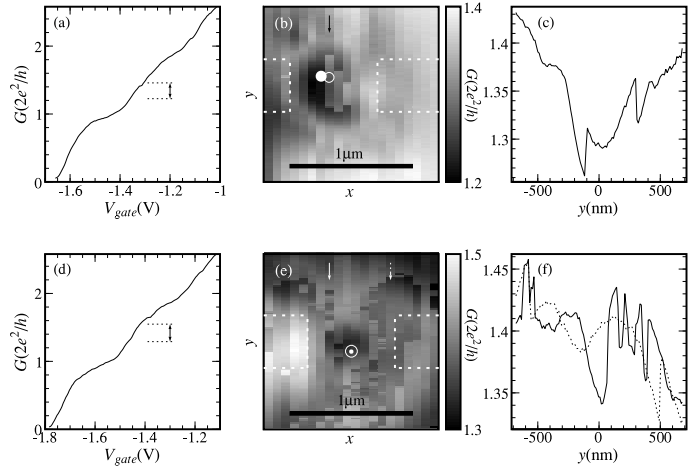}
\end{center}
\caption{Gate sweeps are given in (a) and (d) where the
 conductance range of the corresponding conductance images (b) and (e) are indicated by
 the dotted lines.
The arrows on the conductance images indicate the y direction single sweeps shown in (c) and (f).  
\label{figthree}}
\end{figure}

Switching between stable states in semiconductor devices is frequently observed as a function of time and is known
 as a Random Telegraph Signal (RTS).
When the time constant of the measurement is longer than the switching period a time average
 of the RTS is measured,
 which can cause a small `plateau feature' seen in conductance during gate bias sweeps of
 1D channels~\cite{Cobden91,Cobden92}.
A RTS observed in channel conduction is believed to be caused by the
 probabilistic occupation of a nearby defect state by an electron originating in the 2DEG,
 which then modifies the channel conduction by electrostatics.
Fig.~\ref{figthree} (a) and (d) show gate sweeps where the dotted lines indicate the conductance range of the
 corresponding conductance images (b) and (e).
Within the conductance range a small `plateau' can be seen on both of the gate sweeps,
 characteristic of switching.
The position of the AFM tip also affects the defect state occupation which is observed in the conductance images
 as two levels with a sharp transition.
Two levels are visible in (b) as a brighter region in the centre,
 and in (e) as two darker regions around the ends of the gate electrodes.
Fig.~\ref{figthree} (c) and (f) show single y direction sweeps taken from the corresponding conductance images where they are
 indicated by arrows.
The switch step size is constant over each image at $3.8~\mu$S for (b),
 which would require a change of 9.3mV in $V_{gate}$ to produce an equivalent effect.
We estimate that a change in gate bias corresponds to 13.3~meV/V change in channel potential~\cite{Thomas95},
 which gives an energy change of $120~\mu$eV due to the switch.
From the conductance image we know the defect is located approximately 100~nm from the channel centre,
 and probably in the donor layer between 40~nm and 80~nm above the 2DEG plane.
If the occupation of the defect state is modelled as a point electron source,
 then to provide the correct energy change the electron is required to be $1~\mu$m from the channel,
 and although screening has been ignored, the model is poor.

\par
We propose that the source of the switch is a single electron hopping between two nearby defect states.
Possible locations for these defects are indicated on the conductance image by $\circ$ and $\bullet$,
 where $\circ$ is occupied to produce the background level.
The defects are at positions $z_{\circ}=60$~nm and $z_{\bullet}=64$~nm above the 2DEG plane,
 and are separated by 10.5~nm which is of the order of the average dopant spacing at 4.3~nm
 and near the effective Bohr radius for GaAs.
\par
The two dark switched regions of (e) have the same energy change which suggests that they originate from the
 same defect system.
When the tip is positioned over the channel, and the nearby 2DEG,
 the switch remains in the background state.
This suggests that this defect system
 is physically below the channel where electrons in the channel and 2DEG provide screening.
Again two defect states are required to provide the correct energy change of about $100~\mu$eV,
 at possible $z_{\circ}=-110$~nm and $z_{\bullet}=-100$~nm,
 and indicated on the image with $\odot$ as
 $x_{\circ}=x_{\bullet}$ and $y_{\circ}=y_{\bullet}$.
When the tip is positioned near the gate electrode ends where the 2DEG is depleted,
 the tip potential can penetrate behind the channel,
 raising the potential of state $\circ$ relative to state $\bullet$,
 as state $\bullet$ is closer to the channel and is more effectively screened.
Further from the channel,
 but where the 2DEG is still depleted,
 the tip has less effect and state $\circ$ is again occupied.

\subsection{Images of channel movement due to asymmetric gate bias}
\begin{figure}
\begin{center}
\includegraphics[width=9cm]{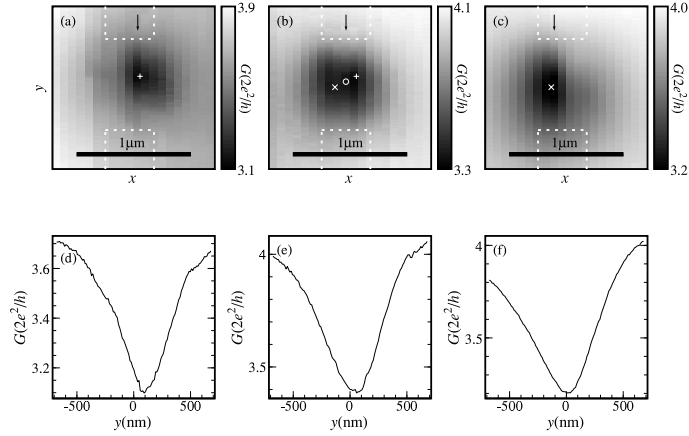}
\end{center}
\caption{The conductance images (a), (b), and (c) were made with the asymmetrically biased gate
 electrodes $(V_{upper},V_{lower})$ of ($-1.36$~V, $-4.91$~V), ($-2.92$~V, $-2.92$~V), and ($-5.19$~V, $-0.92$~V),
 with the channel centre $\left( \Delta x, \Delta y \right)$ relative to $\circ$ measured
 at (98~nm, 52~nm), (0~nm, 0~nm), and ($-103$~nm, $-50$~nm) respectively.
Single sweeps from the conduction images, where they are indicated by arrows,
 are reproduced in (d), (e), and (f).
\label{figfour}}
\end{figure}

For the previous experiments both gate electrodes were biased to the same potential.
By asymmetrically biasing the surface electrodes 
 we observe movement of the channel centre in conductance images.
Channel lateral position has been studied theoretically~\cite{Glazman91},
 and a total movement of $\Delta x = 100$~nm
 has been deduced by the effect of a 
 defect within the channel~\cite{Williamson90},
 though this enabled the channel centre position to be measured in only one direction.
For this experiment we used a slightly wider 800~nm split gate orientated in the y direction,
 but on the same device and with the same AFM tip.
Fig.~\ref{figfour} shows three conductance images made with asymmetric gate biases
 $\Delta V_{gate} = V_{upper} - V_{lower}$ of 3.55~V, 0~V, and $-4.27$~V respectively,
 with average gate biases $\overline{V}_{gate} = (V_{upper} + V_{lower})/2$
 of $-3.14$~V, $-2.92$~V, and $-3.06$~V set to obtain the same initial conductance.
Note that these images were produced consecutively in the order (b), (c), then (a) to avoid
 the possibility of misinterpreting mechanical drift as channel movement.
The channel centre was determined in both the x and y directions by fitting a parabola to
 the region around the conductance minima.
Fig.~\ref{figfour} (d), (e), and (f) show y direction single sweeps reproduced from the 
 respective conductance images where they are indicated by arrows,
 and were selected to include the channel centre in the x direction.
As is evident from the single sweeps the perturbing potential is itself not symmetric,
 which is why a fit to $V_{tip}/\sqrt{\rho^2+(d+r_{tip})^2}$ was rejected,
 and is probably caused by screening of the tip potential by the surface electrodes.
The channel centres are indicated on the conductance images by
 $+$, $\circ$, and $\times$ at positions relative to $\circ$ of
 (98~nm, 52~nm), (0~nm, 0~nm), and ($-103$~nm, $-50$~nm).
The unexpected channel movement in the x direction is caused by
 imperfections in the surface electrodes, or disorder from the doped layer,
 which lead to imperfections in the confining potential.
The effect is principally seen in the x direction due to the much smaller
 x component of the confining potential gradient,
 and therefore increased sensitively to confining potential imperfections.

\section{Conclusion}
We have used a conductive AFM tip to modify the conductance of a 1D ballistic channel.
By modelling the tip as a spherical charge we obtain a good fit to the observed tip potential perturbation.
We have made measurements which reveal structure across the width of the channel,
 corresponding to the predicted charge density for one, two, and three sub-bands.
We have produced images with two levels due to an electron hopping between defect states.
When the defect system was beneath the 2DEG plane we obtained images revealing the 1D channel which screened
 the tip potential.
We have observed channel movement when the surface electrodes were biased asymmetrically.
Movement along the length of the channel is believed to be due to electrode imperfections.

\section*{Acknowledgments}
This work was funded by the EPSRC and the RW Paul Instrument Fund.


\end{document}